%% file: main.tex
\documentclass{amsart}  

\include{misc/config}

\usepackage{eso-pic}
\newcommand\BackgroundPic{%
\put(127,513){%
\centering
\includegraphics[width=\textwidth]{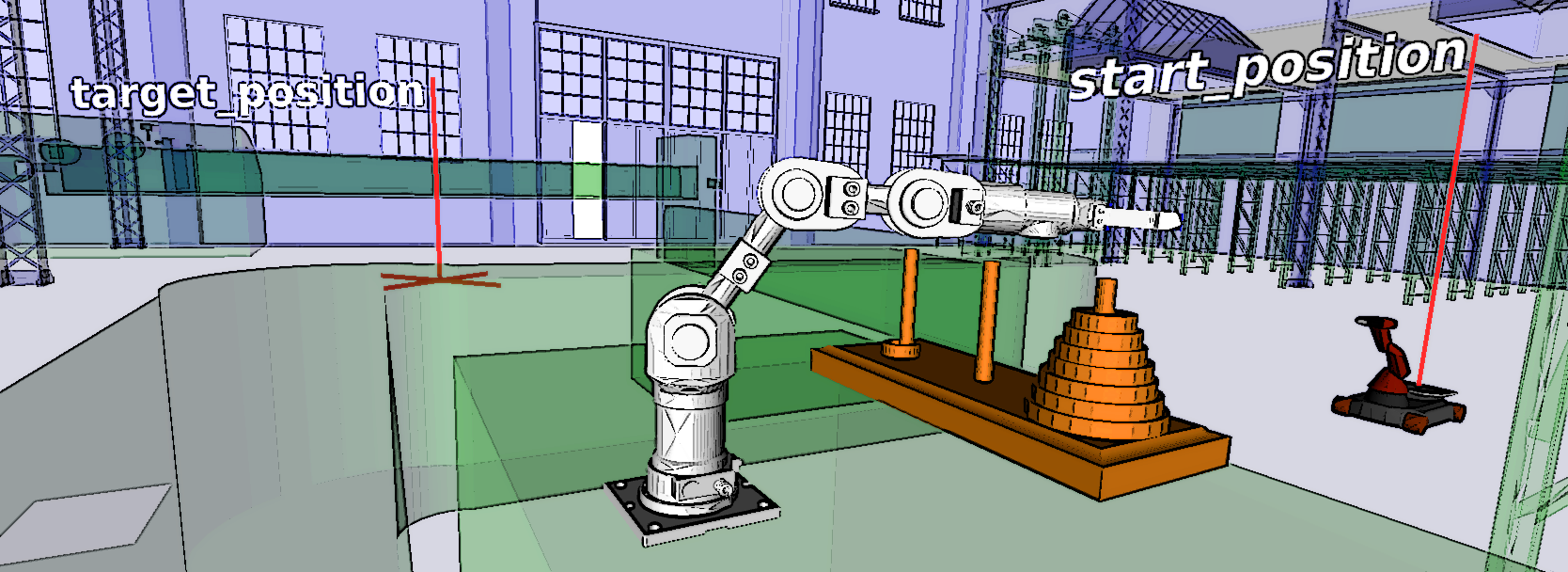}%
}}

\begin{document}
\AddToShipoutPicture*{\BackgroundPic}

\title[Reasoning in complex environments with the \textsc{SelectScript}] 
{.\\[5.5cm] Reasoning in complex environments with\\ the \textsc{SelectScript} declarative language}

\author{Andr\'e Dietrich, Sebastian Zug, Luigi Nardi, and J\"org Kaiser}
\address[Andr\'e Dietrich, Sebastian Zug, and J\"org Kaiser]{%
\Institute,
\indent\University,
Magdeburg, Germany}
\email[]{\{\xEmail{dietrich}{IROS2015}{dietrich},
       \xEmail{zug}{IROS2015}{zug},
       \xEmail{kaiser}{IROS2015}{kaiser}\}@ivs.cs.uni-magdeburg.de}

\address[Luigi Nardi]{%
Department of Computing, Imperial College, London, United Kingdom}
\email[]{\href{mailto:l.nardi@imperial.ac.uk}{l.nardi@imperial.ac.uk}}
\date{\today}


\maketitle

\begin{abstract}
\textsc{SelectScript} is an extendable, adaptable, and declarative
domain-specific language aimed at information retrieval from simulation
environments and robotic world models in an \acs{SQL}-like manner.
In this work we have extended the language in two directions. First, we have
implemented hierarchical queries; second, we improve efficiency enabling manual
design space exploration on different ``search'' strategies. As depicted in the
teaser above, we demonstrate the applicability of such extensions in two
application problems; the basic language concepts are explained by solving the
classical problem of the Towers of Hanoi and then a common path planning problem
in a complex 3D environment is implemented.
\end{abstract}

\input{misc/acronyms}
\input{sec/intro}
\input{sec/stoa}
\input{sec/language}

\input{sec/concept}
\input{sec/application}
\input{sec/outlook}

\section*{Acknowledgments} 
We acknowledge funding by the EPSRC grant PAMELA EP/K008730/1.

\hypersetup{hidelinks=false}
\bibliographystyle{IEEEtran}
\bibliography{bib/literatur,bib/ess}

\end{document}

%% file: misc/config.tex
\usepackage[USenglish]{babel}
\usepackage[utf8]{inputenc}

\usepackage{graphicx}
\usepackage{subfig}
\usepackage{pgf}
\usepackage{pgfplots}
\usepackage{todonotes}

\usepackage{listings}
\lstdefinelanguage{XML}
{
	morestring=[b]",
	morecomment=[s]{<?}{?>},
	morecomment=[s][\color{gray}]{<!--}{-->},
	keywordstyle=\color{cyan},
	tagstyle=\color{black},
	morekeywords={xmlns,version},
	stringstyle=\color{red},
	identifierstyle=\color{blue}
}

\usepackage[nolist]{acronym}

\usepackage[binary-units=true]{siunitx}		
\usepackage{color}

\usepackage{tikz}
\usepackage{tikz-3dplot}
\usepackage{pgffor}

\usetikzlibrary{arrows}
\usetikzlibrary{automata}

\usetikzlibrary{backgrounds}

\usetikzlibrary{calc}
\usetikzlibrary{calendar}
\usetikzlibrary{chains}

\usetikzlibrary{decorations}
\usetikzlibrary{decorations.footprints}
\usetikzlibrary{decorations.fractals}
\usetikzlibrary{decorations.markings}
\usetikzlibrary{decorations.pathmorphing} 
\usetikzlibrary{decorations.pathreplacing}
\usetikzlibrary{decorations.shapes}
\usetikzlibrary{decorations.text}

\usetikzlibrary{er}

\usetikzlibrary{fadings}
\usetikzlibrary{fit}
\usetikzlibrary{folding}

\usetikzlibrary{matrix}
\usetikzlibrary{mindmap}

\usetikzlibrary{patterns}
\usetikzlibrary{petri}
\usetikzlibrary{plotmarks}
\usetikzlibrary{positioning}

\usetikzlibrary{scopes}
\usetikzlibrary{shadows}
\usetikzlibrary{shapes}
\usetikzlibrary{shapes.arrows}
\usetikzlibrary{shapes.callouts}
\usetikzlibrary{shapes.gates.logic.US}
\usetikzlibrary{shapes.gates.logic.IEC}
\usetikzlibrary{shapes.geometric}
\usetikzlibrary{shapes.misc}
\usetikzlibrary{shapes.multipart}
\usetikzlibrary{shapes.symbols}

\usetikzlibrary{through}
\usetikzlibrary{topaths}
\usetikzlibrary{trees}

\usepackage{multirow}
\usepackage{tabularx}
\usepackage{booktabs}

\usepackage{algorithmic}
\usepackage{algorithm}

\usepackage{url}
\usepackage{hyperref}
\hypersetup{
pdfview={FitH \hypercalcbp{\paperheight-\topmargin-1in-\headheight}},
pdfstartview={FitH \hypercalcbp{\paperheight-\topmargin-1in-\headheight}},
pdftitle={Titel}
}

\newcommand{\fig}[1]{Fig.~\ref{#1}}

\newcommand{\lis}[1]{Lis.~\ref{#1}}
\renewcommand{\sec}[1]{Sec.~\ref{#1}}

\usepackage{pdfsync}

\newcommand{\eg}[0]{e.\,g.}
\newcommand{\ie}[0]{i.\,e.}

\newcommand{\University}[0]{%
\href{http://www.uni-magdeburg.de}{Otto-von-Guericke-Universit\"at Magdeburg}}

\newcommand{\Institute}[0]{%
\href{http://ivs.cs.uni-magdeburg.de}{Department of Distributed Systems}}

\newcommand{\mail}{ivs.cs.uni-magdeburg.de}

\newcommand{\xEmail}[3]{\href{mailto:#2@\mail?subject=#3}{#1}}

%% file: misc/acronyms.tex
\begin{acronym}
\acro{AmC}           {Ambient Computing}
\acro{AmI}           {Ambient Intelligence}
\acro{API}           {Application Programming Interface}
\acro{ARToolkit}     {Augmented Reality Toolkit}
\acro{AWDS}          {Ad-hoc Wireless Distribution Service}
\acro{CAD}           {Computer-Aided Design}
\acro{CAMUS}         {Context-Aware Middleware for \acs{URC} System}
\acro{CAN}           {Controller Area Network}
\acro{COLLADA}       {COLLAborative Design Activity}
\acro{CORBA}         {Common Object Request Broker Architecture}
\acro{CoTeSys}       {Cognition for Technical Systems}
\acro{CPS}           {Cyber Physical Systems}
\acro{DAvinCi}       {Distributed Agents with Collective Intelligence}  
\acro{DBMS}          {Database Management System}
\acro{DDL}           {Device Description Language}
\acro{DSL}           {Domain-Specific Language}
\acro{ECA}           {Event Condition Action}
\acro{EM}            {Environment Model}
\acro{FAMOUSO}       {Family of Adaptive Middleware for autonomOUs Sentient Objects}
\acro{GIS}           {Geographic Information System}
\acro{GOLOG}         {alGOl in LOGic}
\acro{GUI}           {Graphical User Interface}
\acro{HCI}           {Human Computer Interaction}
\acro{HDFS}          {Hadoop Distributed File System}
\acro{ID}            {IDentifier}
\acro{IE}            {Intelligent Environments}
\acro{IEEE}          {Institute of Electrical and Electronics Engineers}
\acro{IoT}           {Internet of Things}
\acro{IoS}           {Internet of Services}
\acro{IS}            {Information Science}
\acro{JDL}           {Joint Directors of Laboratories}
\acro{JVM}           {Java Virtual Machine}
\acro{LDM}           {Local Dynamic Maps}
\acro{LISP}          {LISt Processing}
\acro{LOP}           {Language-Oriented Programming}
\acro{M2M}           {Machine-to-machine communication}
\acro{MavHome}       {Managing An Intelligent Versatile Home}
\acro{MMT}           {Mental Model Theory}
\acro{MOSAIC}        {fraMewOrk for fault-tolerant Sensor dAta fusIon in dynamiC environements}
\acro{NASA}          {National Aeronautics and Space Administration}
\acro{NCAP}          {Network Capable Application Processor}
\acro{nesC}          {network embedded systems C}
\acro{NI}            {Network Interface}
\acro{NoSQL}         {Not only SQL}
\acro{NRS}           {Network Robot System}
\acro{ODE}           {Open Dynamics Engine}
\acro{OGC}           {Open Geospatial Consortium}
\acro{OMG}           {Object Management Group}
\acro{OpenGL}        {Open Graphics Library}
\acro{OpenRAVE}      {OPEN Robotics Automation Virtual Environment}
\acro{OSGi}          {Open Services Gateway initiative}
\acro{OWL}           {Web Ontology Language}
\acro{PCT}           {Perceptional Control Theory}
\acro{PEIS}          {Physically Embedded Intelligent System}
\acro{PerComp}       {Pervasive Computing}
\acro{PLUE}          {Programming Language for Ubiquitous Environments}
\acro{PREDiMAP}      {vehicle Perception and Reasoning Enhanced with Digital MAP}
\acro{Prolog}        {PROgrammation en LOGique}
\acro{RDF}           {Resource Description Framework}
\acro{RDQL}          {Resource Description Query Language}
\acro{RFID}          {Radio-Frequency IDentification}
\acro{ROS}           {Robot Operating System}
\acro{RPC}           {Remote Procedure Call}    
\acro{SensorML}      {Sensor Model Language}
\acro{SLAM}          {Simultaneous Localization And Mapping}
\acro{SmE}           {Smart Environments}
\acro{SOA}           {Service Oriented Architecture}
\acro{SQL}           {Structured Querying Language}
\acro{STRIPS}        {Stanford Research Institute Problem Solver}
\acro{SWE}           {Sensor Web Enablement}
\acro{TCP}           {Transmission Control Protocol}
\acro{TEDS}          {Transducer Electronic Data Sheet}
\acro{tf}            {transformation}
\acro{TII}           {Transducer Independent Interface}
\acro{TIM}           {Transducer Interface Module}
\acro{TransducerML}  {Transducer Markup Language}
\acro{TTL}           {Time To Live}
\acro{UAV}           {Unmanned, Uninhabited, or Unpiloted Aerial Vehicle}
\acro{UbiBot}        {Ubiquitous Robotics}
\acro{UbiComp}       {Ubiquitous Computing}
\acro{UDP}           {User Datagram Protocol}
\acro{UML}           {Unified Modeling Language}
\acro{URC}           {Ubiquitous Robotic Companion}
\acro{URDF}          {Unified Robot Definition Format}
\acro{URL}           {Uniform Resource Locator}
\acro{VTK}           {Visualization Toolkit}
\acro{XML}           {eXtensible Markup Language}
\end{acronym}

%% file: sec/intro.tex
\section{Introduction}

We have developed \textsc{SelectScript} \cite{dietrich2015icra} as a consequence
to the growing complexity of (robotic) world models and discrete simulation
environments that we are confronted with in our everyday scientific life. If we
look at the world of robotics, smart environments, or cyber-physical systems,
all of the entities within such newly developing ecologic systems posses their
own simulation of the environment. Each of them with a very specific structures
and \acsp{API} and different semantics. An intermediate language that can be put
on top of all these world models with a common semantic, would enable a ``real''
interoperability between these systems (not in the sense of sharing data, data
sheets, or services). An entity could access the required information from
another entities ``head'', by defining what information it requires and in what
representation format. \textsc{SelectScript} therefore adopts the well known
\acs{SQL} syntax for discrete simulations, allowing to query them online and to
manually explore the capabilities of the \acs{API} together with the underlying
environment. Within the following, we will use the terms simulation and world
models synonymously, since a world model can be interpreted as a simulation of
certain aspects of the world, further more we are demonstrating the capabilities
of our query language by applying different simulation environments.

\textsc{SelectScript} is a declarative domain-specific language (\acs{DSL})
currently embedded in Python, further language implementations are planned.
It is called \textsc{SelectScript} because, in contrast to a complete \acs{SQL}
implementation, we only provide the possibility for accessing and querying data,
which is based  on \texttt{SELECT}-statements, the development and integration
of additional functionality  is left out to the host programming  language, \eg\
Python, C++, etc. Additionally, we added new language features  that  allow to
query for application-specific response formats, \eg\ maps, sound, projections,
etc., integrate temporal aspects, and offer interfaces that enable to extend the
language according to different requirements (cf. \sec{sec:language}).

In this work we extend the \textsc{SelectScript} \acs{DSL} to enable
hierarchical queries. This does not only allow to solve classical reasoning
problems such as the Towers of Hanoi or the Four Color Map problem, but also
path planning in a complex 3D environment, as indicated by the teaser figure.
Seeking for an appropriate method to resolve such hierarchical queries by
implementing different strategies, we shortly realized that is actually possible
and even more valuable to integrate different ones. But this requires to
implement mechanisms that allow to switch between different strategies and to
tweak them. Strategies with their specific details of implementation, memory
management, etc. can thus be abstracted and hidden behind the language. The
\textsc{SelectScript} \acs{DSL} allows to define what information is required as
well as what ``search'' strategy to apply. No one would normally apply an
\acs{SQL}-like syntax to solve common reasoning problems, whereby it actually
provides a convenient way of expressing problems.\\

\noindent The contributions of this paper are:

\begin{itemize} 
\item We extend the \textsc{SelectScript} \acs{DSL} with hierarchical queries, 
which enable recursion within the language. 
\item We enable manual design space exploration on different ``search''
strategies.
\item We demonstrate the language improvements by implementing the Towers of
Hanoi application and we show that the same concepts can be used in complex 3D
environments to solve a common path planning problem. 
\end{itemize}

\subsection*{Overview}

The next section is used to give a brief overview on related approaches.
It is followed by an introduction into the basic and original language concept
of \textsc{SelectScript}. In \sec{sec:concept} we discuss why a search problem
can, in general, be seen as a table containing all possible intermediate steps,
such that it becomes apparent that a declarative approach provides a convenient
way to identify the relevant ``rows'' and thus solutions. The same section then
introduces hierarchical queries. As already mentioned, the Towers of Hanoi are
used to demonstrate stepwise the new concepts behind this work in
\sec{sec:towers} and a path planning problem in a complex 3D environment is then
solved in \sec{sec:path_planning}. A short summary is presented and an outlook
onto future directions provided.

\enlargethispage{-\baselineskip}

%% file: sec/stoa.tex
\section{Related Work}
\label{sec:stoa}
\enlargethispage{\baselineskip}

This work is an extension of a previous work from the same authors, an
interested reader can find \textsc{SelectScript}\footnote{A simplified version
of the grammar and the base implementation can be found at:\\
$ $ \hfill \url{https://gitlab.com/OvGU-ESS/SelectScript}} explained in
\cite{dietrich2015icra}. In that work an example of the OPEN robotics automation
virtual environment (\acs{OpenRAVE}) \cite{diankov_thesis} is provided.
Nonetheless, we briefly introduce the basics of the \textsc{SelectScript}
language in \sec{sec:language}.

DSLs have been widely used in research and industry to improve expressiveness
for a variety of application domains such as robotics, computational science,
computational finance, image processing, music, graphics, artificial
intelligence \cite{proceeding:LuigiNardiYAO2009,Rathgeber2015,
orlarey2009faust,ragan2013halide}. As an example, in
\cite{proceeding:LuigiNardiYAO2009} the user is able to represent his
variational data assimilation application under a specifically designed
formalism. The compiler automatically generates code improving user
productivity. The functionalities provided allow a user to run simulations in a
specific world model representation, namely a structured mesh. A similar
approach on unstructured meshes is introduced by \cite{Rathgeber2015} where a
high-level specification is provided using a syntax close to the mathematics,
\ie\ the partial differential equations, the software is intended to solve.
The \textsc{SelectScript} DSL differs from these examples in the fact that it
targets a general interface to world models and simulations.

Another class of DSL is given by languages that are performance-oriented. These
languages are often called performance \acsp{DSL} \cite{LuigiNardiHPCC2012,
Rathgeber2012, proceeding:PLuTo}. As an example \cite{Rathgeber2012} implements
an access-execute model where the user specifies a stencil computation for
unstructured meshes; the \acs{DSL} decides how to execute the computation taking
into account manual and automatic performance optimisation techniques.
Effective manual design space exploration is also proposed by
\cite{ragan2013halide} where the authors provide a mechanism for manually
exploring different schedule choices. As we shall see, \textsc{SelectScript}
also proposes an effective manual design space exploration.

The robotics community has been active on the creation of \acsp{DSL} for
robotics. A compelling example that shares some commonalities with our work is
RSG-DSL \cite{blumenthal2014towards}, which is a \acs{DSL} focusing on robot
scene graphs (RSGs) world model representations. The restriction of considering
RSGs-only is overcome by \textsc{SelectScript}. Another example is the robot
perception specification language (RPSL) \cite{hochgeschwender2014towards} which
focuses on task knowledge and the variety of sensors a robot has to deal with
when implementing a robot perception architecture~(RPA). MeshSQL
\cite{lee2004meshsql} defines queries for mesh-based physics simulations.
In contrast to our approach, the results of a simulation are stored within a
database, according to time and space. MeshSQL is thus a real extension of
SQL1999 and specialized on mesh data only. This query language is intended to
enable researchers to interactively explore simulation data, to identify new and
interesting effects. MeshSQL therefore offers temporal, spatial, statistical,
and similarity queries, which require different types of return values, \ie\
simple values, surfaces and slices in different formats. Other approaches adopt
the \acs{SQL} syntax. For example the language integrated query (LINQ)
\cite{pialorsi2007linq} extends some of the .NET languages to apply \acs{SQL}
queries to relational databases as well as to arrays. This is similar to other
approaches in the Java world, Java query language (JQL)
\cite{willis2006efficient}, SQL for Java Objects (JoSQL) \cite{webJoSQL}, etc.,
but with a reduced syntax. All these approaches are a close match with the
actual \acs{SQL} language and do not meet the requirements of simulations and
robotic world models nor the requirements of hierarchical querying.

%% file: sec/language.tex
\section{Language Basics}
\label{sec:language}

\begin{figure}[thb]
\centering
\subfloat[$ t_{sim} = 10$]
{\includegraphics[width=0.5\linewidth]{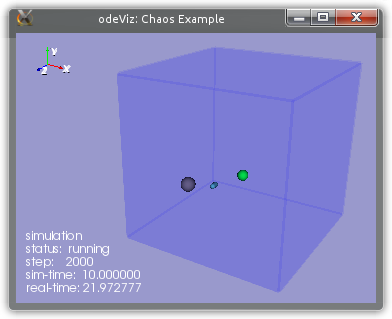}}\hfill
\subfloat[$ t_{sim} = 100$]
{\includegraphics[width=0.5\linewidth]{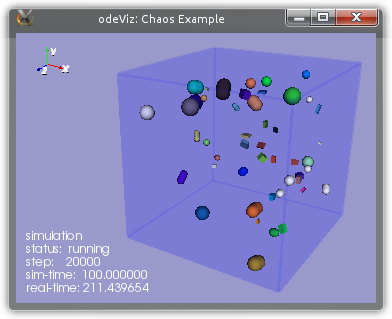}}\qquad
\subfloat[$ t_{sim} = 250$]
{\includegraphics[width=\linewidth]{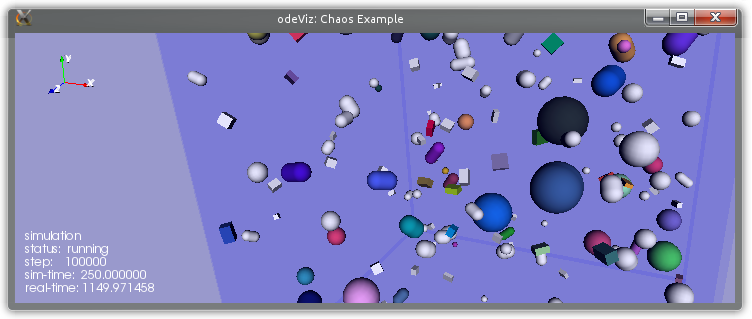}
\label{fig:chaos3}}\qquad
\caption{Screenshots of the particle simulation at different time steps.}
\label{fig:chaos}
\end{figure}

We will use the interpreter for the open dynamics engine (\acs{ODE})
\cite{web:ode} on the ``chaotic'' simulation\footnote{Demo:
\url{http://www.youtube.com/watch?v=F1XNch1JC9Y}} of \fig{fig:chaos} to
demonstrate the \textsc{SeleceScript} basics and to show that the interpreter
can be used on top of multiple environments. All the presented examples within
this paper as well as the language implementation can be downloaded via
git\footnotemark[1].\\[1mm]
Additionally, the commented slides with the complete code can be downloaded as
an IPython notebook at:
{\small\url{http://nbviewer.ipython.org/url/gitlab.com/OvGU-ESS/SelectScript_demos/raw/master/DSLRob-15/presentation.ipynb}}\\[2mm]
Furthermore, there are multiple different screencasts available at our
YouTube-channel at: {\small\url{http://www.youtube.com/ivsmagdeburg}}\\[1mm]
\indent The example in \fig{fig:chaos} depicts a chaotic but configurable
simulation\footnote{Source: \url{https://gitlab.com/OvGU-ESS/odeViz}} in
\acs{ODE}. Objects of different size, shape, color, mass, direction, and speed
appear within a box. These particles fly around, bump against each other, and
colliding objects can explode, giving birth to other ``white particles'', or
vanish. The more time passes by, the more objects will be present in the box.
All of the object parameters and their place of appearance are random values.

\begin{figure}[h!]
\centering
\includegraphics[width=\linewidth]{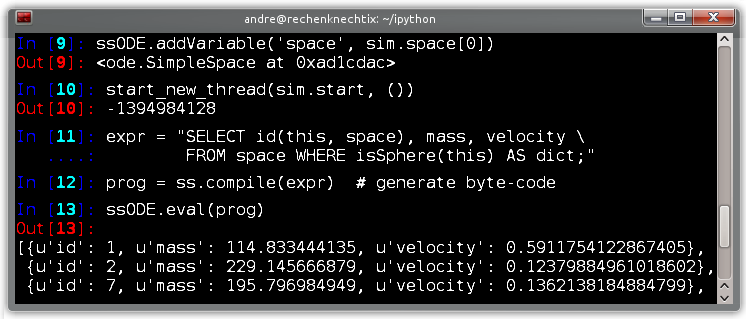}
\caption{Screenshot of interactive data exploration via ipython.}
\label{fig:dataminig}
\end{figure}

How would you analyse this? In a traditional way we would use the \acs{API} of
the simulation framework and a programming language with a lot of loops. Instead
\fig{fig:dataminig} shows how such analysis can be led using
\textsc{SelectScript} interactively. A user has to attach the \acs{ODE} space
variable, which is the host of the entire simulation environment, to the
interpreter to make it accessible under the new variable name \texttt{"space"}.
The \textsc{SelectScript} interpreter for \acs{ODE} checks the variable type,
\eg\ whether it is a list, a dictionary, or an \acs{ODE} space, etc., and
applies the appropriate methods.

As further depicted in \fig{fig:dataminig} and in \fig{fig:exploration}, the
\texttt{"space"} variable can then be accessed in the \texttt{FROM}, as if it
would be an ordinary table. As further visible, the listed ``columns'' within
the \texttt{SELECT} expression are actually function calls, which offer a
convenient way to abstract the \acs{API}. All functions, with the exception of
some internals such as \texttt{eval}, \texttt{help}, \texttt{print}, or
\texttt{to}, are implemented externally in the host programming language and can
be applied everywhere within a script. Functions are linked to the
\textsc{SelectScript} interpreter similarly as it was done for variables, see
also \lis{lis:python} on page \pageref{lis:python}. An additional element, that
does not belong to the standard \acs{SQL} syntax, is the keyword \texttt{this},
which is a pointer pointing to the currently evaluated element in the
\texttt{FROM} expression. Since functions can have multiple input parameters, it
allows to call them with the correctly placed parameter sets, see the function
\texttt{id} in the figure above. If a function is called within the
\texttt{SELECT} expression and if it has only one input parameter, the
\texttt{this} pointer is automatically passed without the need to write it in
brackets, as for the functions \texttt{mass} and \texttt{velocity} in
\fig{fig:dataminig}. Everywhere else within a script functions are identified by
a function name with following brackets. As already mentioned and depicted in
\fig{fig:dataminig}, \textsc{SelectScript} is a language embedded in Python.
The source code of a script is compiled at runtime into an intermediate
representation and then evaluated. An interaction between Python and our query
language is described in more detail in the application of \sec{sec:towers}.


A \textsc{SeleceScript} script can contain multiple queries, as depicted for the
listing in \fig{fig:exploration}, and queries can be nested. Intermediate
results can be stored persistently in variables, allowing to reuse results and
to define more complex query scripts. Thereby the last statement of a script is
defining the return value and can therefore also be a composition of multiple
results, as for example in line $7$ of \lis{lis:select-abstactions}.
Additionally \textsc{SelectScript} has also support for temporal variables,
which allows previous results to be cached over a certain period of time.

\noindent\begin{minipage}{\linewidth}
  \vspace{4mm}
  \centering
  \begin{minipage}[t]{.37\linewidth}
  \centering
   \begin{lstlisting} %
[label=lis:select-examples, %
keywords={SELECT, FROM, WHERE, AS}, %
keywordstyle=\color{red}\ttfamily\scriptsize]
 # SelectScript-Example
 spheres =
   SELECT obj
   FROM space
   WHERE isSphere(this)
   AS list; 
 maxMass = max(
   SELECT mass
   FROM spheres
   AS list);
\end{lstlisting} 
  \end{minipage}\hspace{9mm}
  \begin{minipage}[t]{.34\linewidth}\hspace{1.2mm}
    \includegraphics[width=\linewidth]{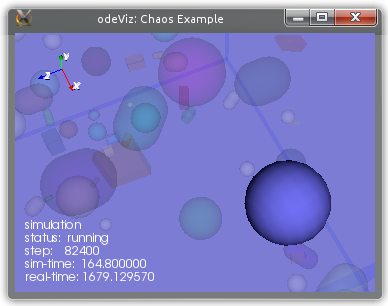}
    \end{minipage}
  \captionof{figure}{\textsc{SelectScript} example for identifying the heaviest
  sphere (left) and visualization with reduced opacity for complementing
  objects (right).}
  \label{fig:exploration}
\end{minipage}\\

\begin{figure}[htb]
\centering
\subfloat[masses (heat map)]
{\includegraphics[width=0.45\linewidth]{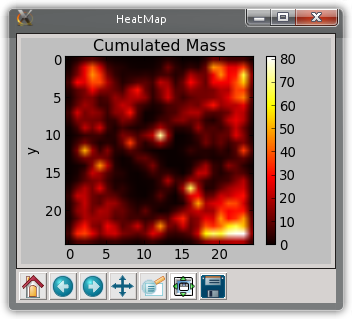}}\hfill
\subfloat[z-velocities (spectral map)]
{\includegraphics[width=0.45\linewidth]{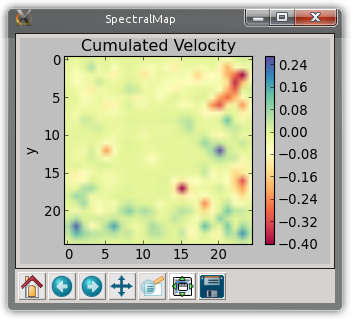}}
\caption{Two 2D projections showing the results of the script in
\lis{lis:select-abstactions}, for cumulated masses and velocities.}
\label{fig:abstractions}
\end{figure}

Next to the basic \texttt{SELECT} \ldots \texttt{FROM} \ldots \texttt{WHERE}
syntax it is also possible to use the common \texttt{GROUP BY} and \texttt{ORDER
BY}, but it is also possible to request certain response format. As it is
depicted in the previous examples, the keyword \texttt{AS} followed by
\texttt{value}, \texttt{list}, or \texttt{dict}, allows a representation format
to be chosen. It has to be noted that if a dictionary is requested the function
names within the \texttt{SELECT} expression are used to define dictionary
\acsp{ID}. But it is also possible to implement new methods, see therefore the
example in \lis{lis:select-abstactions}. The requested results (\texttt{plane})
are projection matrices that can be further parameterized. The resulting
2D-matrices for mass and velocity that are returned as elements of a list, are
afterwards visualized as a heat map and a spectral map (using Python
Matplotlib\footnote{\url{http://matplotlib.org}}), see \fig{fig:abstractions}.
In \cite{dietrich2015icra} it was further demonstrated for \acs{OpenRAVE}, that
it is possible to request a result \texttt{AS OccupancyGrid} for external
planners, \texttt{AS Prolog} to support reasoning tasks, or even \texttt{AS
SensorMap} that depicts the current sensor coverage of an area.

\begin{lstlisting} %
[label=lis:select-abstactions, %
keywords={SELECT, FROM, WHERE, AS}, %
keywordstyle=\color{red}\ttfamily\scriptsize,%
caption={\textsc{SelectScript} with a requested visual abstractions of the
simulation. The additional parameters for the plane projection are: the plane
itself (1st), the starting point (2nd), the width, the height, and the
resolution (3rd), and an optional blur (4th) parameter.}]
 # Select 2 dimensional projections ...
 Mass    = SELECT mass FROM space WHERE hasBody(this) 
           AS plane('XY', [-5,-5], [100,100,0.1], 3);
 Velocity= SELECT linearVelocity(this, 2) FROM space
           WHERE hasBody(this)
           AS plane('XY', [-5,-5], [100,100,0.1], 3);
 [Mass, Velocity];                    # return values
\end{lstlisting} 

%% file: sec/concept.tex
\section{DSL Extension}
\label{sec:concept}

\subsection{Graph and Table Representation Equivalence}

As depicted in \fig{fig:tree_vs_table} most of the common search problems can be
represented as either a graph/tree or as a table, containing all possible
combinations and thus paths within the graph or tree. It is obvious that a
graph-like structure consumes less memory than a table, whereby a graph needs to
be traversed every time in order to find an appropriate solution. A table that
contains all ``paths'' can be explored more efficient, whereby different caching
or indexing strategies are applied in the background, and filters can be
applied, \eg\ by directly requesting the row with the last entry $cc$.

\begin{figure}[htb]
    \vspace{-4mm}
    \subfloat[Connected undirected graph] {
        \begin{minipage}[]{5.3cm}
        \centering
        \input{pic/graph.pgf}
        \label{fig:graph}
        \end{minipage}
    }\hspace{5mm}
    \subfloat[Table of poss. paths] {
        \begin{minipage}[]{4.2cm}
        \centering
        \input{pic/table.pgf}
        \label{fig:table}
        \end{minipage}
    }    
    \caption{
    The Tower of Hanoi problem with two disks. Both
    representations can be used to identify a valid path and thus the
    intermediate to get from configuration $aa$ to $cc$. The letters 
    mark the positions of the disks (tower) and their order is used to
    mark their size (starting with the smallest).}
    \label{fig:tree_vs_table}
\end{figure}

Nevertheless, we only apply the database metaphor onto common search and
reasoning problems. The ``graph/tree'' therefore still has to be created and
traversed, but this is handled in the background. Applying an \acs{SQL}-like
syntax offers a more ``natural'' way to express problems and it furthermore
offers a method to utilize different search algorithms and methods for
optimization, without the need for changing the problem description. Note that
the order of rows within the table in \fig{fig:table} is a result of an applied
depth-first search algorithm.

\subsection{Hierarchical Queries}

A hierarchical or recursive query in \acs{SQL} is a special type of query used
to handle hierarchical structured data. A common example is a company database,
which stores all employees data as well as the \acs{ID} of the direct
supervisor. Thus, reconstructing the leadership structure requires additional
capabilities that enable recursion. The \lis{lis:hierarchy} shows an
alternative, not standardized, syntax for recursive queries that we implement in
\textsc{SelectScript}. It adopts the \texttt{START WITH} \ldots\ \texttt{CONNECT
BY} construct that was originally introduced by Oracle
\cite{przymus2010recursive}. The \acs{SQL} standard \cite{eisenberg1999sql}
defines a syntax that involves the keywords \texttt{WITH RECURSIVE} and requires
to associate query expressions with a name, which allows to reuse them. However,
since we are not dealing with tables of finite length but rather with functions
that might generate endless results, we had to integrate an additional
\texttt{STOP WITH} construct, used to define abort conditions. The \texttt{START
WITH} expressions are used, as the name suggests, to define local variables that
might be required and to define initial conditions.

\begin{lstlisting} %
[label=lis:hierarchy, %
firstnumber=1, %
keywords={SELECT, FROM, WHERE, START, WITH, CONNECT, BY, STOP, NO, CYCLE,
UNIQUE, MEMORIZE, MAXIMUM}, %
keywordstyle=\color{red}\ttfamily\scriptsize, %
caption={Generic syntax to support hierarchical queries in
\textsc{SelectScript}.}]% We use ... to depict a missing parameter.}]
 
 SELECT ...  FROM ... WHERE  ...

 START WITH value = ..., ...             # start expressions
 CONNECT BY (NO CYCLE | UNIQUE | MEMORIZE int | MAXIMUM int)
            value = func(value, ...), ...        # recursion
 STOP WITH  value > ... or value ...       # stop conditions
\end{lstlisting}

All of the work is actually done with the help of the central \texttt{CONNECT
BY} expressions. It is used to denote which values within the search are
affected and how they are changed from iteration to iteration, in this case
using a previously defined function. The elements listed above, namely
\texttt{SELECT}, \texttt{FROM}, \texttt{WHERE}, \texttt{START WITH},
\texttt{CONNECT BY}, and \texttt{STOP WITH}, are sufficient to describe the
entire search problem but, as previously mentioned, we also want to include the
possibility to explore different search strategies. The design space exploration
can be manually performed using the accompanying keywords in the \texttt{CONNECT
BY} expression. Each of them tweaks the current, or applies another, search
strategy. The method that is applied at default is a depth-first search
algorithm, which in contrast to the table provided in \fig{fig:tree_vs_table}
would generate more results by allowing cycles (cf. \lis{lis:cycle}). In some
situations cycles within the search result might be desired, but to prevent them
and therefore to reduce the search space, the \texttt{NO CYCLE} keyword has to
be added to the query (cf. \lis{lis:no_cycle}). The application of the keyword
\texttt{UNIQUE} further reduces the space of possible paths, by allowing a node
within the graph to be traversed only once (cf. \lis{lis:no_cycle}). The
application of \texttt{MEMORIZE} automatically generates a connected graph at
first, similar to the one in \fig{fig:tree_vs_table}, which is afterwards
traversed with a bidirectional\footnote{For more information, see our benchmark
at: \url{www.aizac.info/bi-graph-search-benchmark}} search algorithm. This
reduces the search space by half trading memory consumption and prevents
multiple visits of a node, but it requires an additional stop parameter that
determines the maximal length of the resulting paths. This maximal path length
is defined by the associated integer value (cf. \lis{lis:memorize}). The
algorithm has a further advantage, it generates all simple paths, namely a path
with no repeated nodes, ordered by their length, starting with the shortest
path. Finally, the keyword \texttt{MAXIMUM} was introduced to denote the maximum
number of results that is returned, for example $1000$ possible paths for a
mobile robot might be a sufficient result (cf. \lis{lis:openrave}).

%% file: pic/graph.pgf
\begin{tikzpicture}[node distance=1.cm, thick, font=\footnotesize]

  \tikzstyle{every state}=[draw=black,inner sep = 2pt, minimum size=3]

  \node[initial,state] (aa) {$aa$};
  \node[state] (ba) [below right of=aa] {$ba$};
  \node[state] (ca) [below left   of=aa] {$ca$};

  \node[state] (bc) [below right of=ba] {$bc$};
  \node[accepting, state] (cc) [below right of=bc] {$cc$};

  \node[state] (ac) [below left  of=bc] {$ac$};
  \node[state] (cb) [below left  of=ca] {$cb$};
  \node[state] (bb) [below left  of=cb] {$bb$};
  \node[state] (ab) [below right of=cb] {$ab$};

  \draw (aa)  -- (ba);
  \draw (aa)  -- (ca);
  \draw (ba)  -- (ca);
  \draw (ca)  -- (cb);
  \draw (cb)  -- (bb);
  \draw (ab)  -- (bb);
  \draw (cb)  -- (ab);
  \draw (bc)  -- (ac);
  \draw (cc)  -- (ac);
  \draw (ac)  -- (ab);
  \draw (ba)  -- (bc);
  \draw (bc)  -- (cc);
\end{tikzpicture}

%% file: pic/table.pgf
\tikzset{
    table nodes/.style={
        rectangle,
        draw=black,
        align=center,
        minimum height=4.4mm,
        text depth=-0.4ex,
        text height=1.ex,
        inner xsep=0pt,
        outer sep=0pt
    },      
    table/.style={
        matrix of nodes,
        row sep=-\pgflinewidth,
        column sep=-\pgflinewidth,
        nodes={
            table nodes
        },
        execute at empty cell={\node[draw=none]{};}
    }
}

\begin{tikzpicture}[node distance=1.cm, thick, font=\footnotesize]

\matrix (first) at (0,0) [table, text width=7mm, name=table]
{
$aa$ & $ca$ & $cb$ & $bb$ \\
$aa$ & $ca$ & $cb$ & $ab$ \\
$aa$ & $ca$ & $ba$ & $bc$ \\
$aa$ & $ba$ & $ca$ & $cb$ \\
$aa$ & $ba$ & $bc$ & $ac$ \\
$aa$ & $ba$ & $bc$ & \underline{$cc$}  \\
};

\end{tikzpicture}

%% file: sec/application.tex
\section{Appications}
\label{sec:application}

The following two sub-sections are used to demonstrate the applicability of our
approach to solve reasoning problems. Therefore we choose the two examples that
were already depicted in the teaser figure on the first page. These are problems
that are commonly not solved by applying an \acs{SQL}-like syntax, but as we
will demonstrate it offers a convenient way. The Tower of Hanoi \sec{sec:towers}
is used to present the previously introduced new parts of our language step by
step, while \sec{sec:path_planning} is used as a proof of concept that
demonstrates the language is applicable to solve robotic problems in a complex
3D environment.

\subsection{Tower of Hanoi}
\label{sec:towers}

To present all steps that are required to solve the Towers of Hanoi problem, we
start with the basic Python functionality that is required. The code snippet in
\lis{lis:python} is therefore used to load the basic interpreter module (line
$1$). If needed, a \textsc{SelectScript} user is required to implement
additional functions within the host programming language, as it is done at line
$3$ with the \texttt{move} function. This function is used to execute valid
steps, it therefore takes as input the configuration of the \texttt{towers} (a
list of lists) and a \texttt{step} to be performed $[$\texttt{from,\,to}$]$.
The following examples are used to solve the Tower of Hanoi problem for three
disks, which requires at least seven steps. The minimal number of steps can be
calculated with the formula $2^{\text{disks}} - 1$. The list
$[[$\texttt{3,\,2,\,1}$]$\texttt{,}$[$\,$]$\texttt{,}$[$\,$]]$ thus denotes the
initial configuration, where the three towers are represented by three lists and
the first tower has the three ordered disks, namely \texttt{3} large disk,
\texttt{2} medium disk, and \texttt{1} small disk, while
$[[$\,$]$\texttt{,}$[$\,$]$\texttt{,}$[$\texttt{3,\,2,\,1}$]]$ defines the final
configuration. The result of the \texttt{move} function is either a new
configuration of towers or an empty list in case of an invalid step. This
function is made accessible from within \textsc{SelectScript} in line $12$.
All of the following \textsc{SelectScript} listings that present different
solutions to the problem have to be presented as strings, see therefore line
$14$, before they are compiled at runtime (line $16$) and evaluated (line~$17$).
\enlargethispage{-\baselineskip}

\begin{lstlisting} %
[label=lis:python, %
firstnumber=1 , %
%keywords={SELECT, FROM, WHERE}, %
%keywordstyle=\color{red}\ttfamily\scriptsize, %
caption={Basic Python program stub, including all required instructions for the Tower of Hanoi application.}]
 import SelectScript, SelectScript.Interpreter

 def move(step, towers):
     if not towers or not towers[step[0]]: pass
     elif not towers[step[1]] or towers[step[1]][-1]>towers[step[0]][-1]:
         #append element on top from another tower top
         towers[step[1]].append(towers[step[0]].pop())
         return towers
     return []

 ss = SelectScript.Interpreter()
 ss.addFunction('move', move)

 problem = """ insert SelectScript here """
 
 ir      = SelectScript.compile(problem)
 result  = ss.eval(ir)
\end{lstlisting}

The example in \lis{lis:example0} presents a vanilla approach, whereby recursion
is applied with the nesting of seven \texttt{move} functions. That means, that
this script actually iterates through all possible configurations, which results
in a relatively long evaluation time of more than $10$ seconds. We report in the
listing captions the evaluation time of the scripts, so that the performance of
different approaches can be compared. A Intel Core i5-4200U
(\SI{1.6}{\giga\hertz} \SI{3}{\mega\byte}) machine with \SI{8}{\giga\byte} of
RAM is used. The code is running on one core of the machine sequentially.
The sequence of resulting steps is printed in the comment in line $12$.
As introduced in \sec{sec:language}, the \texttt{this} keyword is used as a
pointer to specify where to place the elements from the \texttt{FROM}
expression, if multiple elements are declared, then the \texttt{name.this}
notation is applied to refer to those elements.

\begin{lstlisting} %
[label=lis:example0, %
firstnumber=1 , %
keywords={SELECT, FROM, WHERE, AS}, %
keywordstyle=\color{red}\ttfamily\scriptsize, %
caption={Vanilla approach that uses nested \texttt{move} functions to %
iterate over all possible configurations.\hfill {\small (Runtime: \SI{10.09}{sec}})}] 
 moves = [[0,1], [0,2], [1,0], [1,2], [2,0], [2,1]];
 SELECT m1.this,  m2.this,  m3.this,  m4.this,  m5.this,  m6.this,  m7.this
 FROM   m1=moves, m2=moves, m3=moves, m4=moves, m5=moves, m6=moves, m7=moves
 WHERE  [[],[],[3,2,1]] == move(m7.this,
                             move(m6.this,
                               move(m5.this,
                                 move(m4.this,
                                   move(m3.this,
                                     move(m2.this,
                                       move(m1.this, [[3,2,1],[],[]] )))))))
 AS list;
 # result: [[0,2],[0,1],[2,1],[0,2],[1,0],[1,2],[0,2]]
\end{lstlisting}

The application of the newly integrated constructs simplifies the code
dramatically and also speeds up the execution. The \texttt{WHERE} expression is
still applied to identify the target configuration, whereby new towers are
continuously produced by the \texttt{CONNECT BY} expression, as in
\lis{lis:cycle}. The start configuration is defined in line $4$ and the
generation of new intermediate steps ends if either an invalid step was applied
or the depth of the search exceeds the level of seven. In this way it is also
possible to generate longer results than seven steps, simply by adapting the
stop condition. The evaluation result of the query will then be a list that
contains all possible solutions, namely valid sequences with cycles.

\noindent\begin{minipage}{\linewidth}
\begin{lstlisting} %
[label=lis:cycle, %
firstnumber=2, %
keywords={SELECT, FROM, WHERE, START, WITH, CONNECT, BY, STOP}, %
keywordstyle=\color{red}\ttfamily\scriptsize, %
caption={Application of the basic hierarchical query.
\,\, {\small (\SI{0.89}{sec})}}]
 SELECT this FROM moves WHERE [[],[],[3,2,1]] == move(this, tower) 
 
 START WITH tower = [[3,2,1],[],[]],   level=1
 CONNECT BY tower = move(this, tower), level=level+1
 STOP WITH  level==7 or []==move(this, tower);
\end{lstlisting}
\end{minipage}

The search can be tweaked by additional parameters, as for example in the script
\lis{lis:no_cycle}. Since the \texttt{SELECT} expression is used to generate the
intermediate steps, we have to add additional elements, such as the current
tower configuration or the level. Otherwise, no valid results could be generated
from the applicable steps only. In the result shown in the last line in
\lis{lis:example0}, multiple steps have to be repeated in order to reach the
target configuration. Nevertheless, both optimization strategies perform better
than the previous script.

\begin{lstlisting} %
[label=lis:no_cycle, %
firstnumber=2 , %
keywords={SELECT, FROM, WHERE, START, WITH, CONNECT, BY, STOP, NO, CYCLE}, %
keywordstyle=\color{red}\ttfamily\scriptsize, %
caption={Application of two additional keywords that result~in a reduction of
the search space. {\small (\texttt{NO CYCLE}=\SI{0.32}{sec}, \texttt{UNIQUE}=\SI{0.19}{sec})}}]
 SELECT this, tower FROM moves WHERE [[],[],[3,2,1]] == move(this, tower)

 START WITH tower = [[3,2,1],[],[]], level = 1
 CONNECT BY NO CYCLE # or UNIQUE
            tower = move(this, tower), level = level+1
 STOP WITH  level == 7 or [] == move(mov.this, tower);
\end{lstlisting}

\lis{lis:memorize} demonstrates the application of the
bidirectional\footnotemark[4] graph search method, it translates the entire
query into an equivalent graph that gets analyzed afterwards. This might reduce
evaluation time, but only for larger problems and by sacrificing memory. Since
the additional parameter already defines the maximum resulting path length, it
makes the application of an additional \texttt{level} variable, to control the
length, unnecessary. Note that the problem below now contains four disks, which
require at least $15$ steps.

\begin{lstlisting} %
[label=lis:memorize, %
firstnumber=2 , %
keywords={SELECT, FROM, WHERE, START, WITH, CONNECT, BY, STOP, MEMORIZE}, %
keywordstyle=\color{red}\ttfamily\scriptsize, %
caption=Application of a bidirectional graph search algorithm.
%$ $ \hfill {\small Runtime:~{\SI{0.1735}{sec}}}]
$ $ \hfill {\small ({\SI{0.17}{sec})}}]
 SELECT this, tower FROM moves WHERE [[],[],[4,3,2,1]] == move(this, tower)
 
 START WITH tower = [[4,3,2,1],[],[]] 
 CONNECT BY MEMORIZE 15 
            tower = move(this, tower)
 STOP WITH  [] == move(this, tower);
\end{lstlisting}

\subsection{Robot Navigation}
\label{sec:path_planning}

The following example is used to demonstrate that \textsc{SelectScript} can also
be applied to more complex environments when using the improvement described by
our work. The teaser figure on the first page shows a screenshot of the path
planning task for a mobile YouBot within a complex industrial environment, which
can be solved with the script in \lis{lis:openrave}. A screencast\footnote{Demo:
\url{http://www.youtube.com/watch?v=EFRV0JSdK3M}} showing the running evaluation
was also uploaded to our YouTube-Channel.

The depicted program structure pretty much reflects the previous examples. The
first three lines are used to define relevant parameters, such as the \acs{ID}
of the robot, its target position, and the directions the robot is allowed to
move (at each step). The applied functions \texttt{move}, \texttt{position},
\texttt{distance}, and \texttt{checkCollision} were defined externally. The
\texttt{move} function here really moves the virtual robot in the virtual
environment and \texttt{checkCollision} is used to identify (as its name
suggests) if the object that is passed as a parameter is colliding with
something within the environment or not.

Thus, the only things that might look a bit unfamiliar in this example are the
more complex stop conditions. Within the stop expression the short-cuts of a
collision detection and some background knowledge according to the application
of the \texttt{distance} function are applied. Thus, paths are cut off from a
further search, if the distance from the current position to the target position
cannot be reached anymore with the remaining number of steps. The
\texttt{IF}\footnote{It is an expression, that can be used for logging or even
to adapt the query, and which can be placed everywhere. The last expression in
the then or else block thereby defines the return value. Thus, the previous
result of \texttt{checkCollision} can also be changed to \texttt{False} (cf.
line $18$ in \lis{lis:openrave}).
\\
\texttt{IF(
condition ;}\\
$ $~~~~~~~\texttt{then, execute, a, sequence, of, expression ;}\\
$ $~~~~~~~\texttt{else, another, sequence, of, expressions )}}-condition is an
additional (experimental) language feature, which is applied at this case to
give an acoustic feedback, only if a collision has occurred (listen to the
video\footnotemark[6]).

\begin{figure}[ht]
    \begin{tikzpicture}[>=latex, inner frame sep=-0mm, inner sep=0pt, remember picture]
        \node at (0,0) {\includegraphics[width=\linewidth]{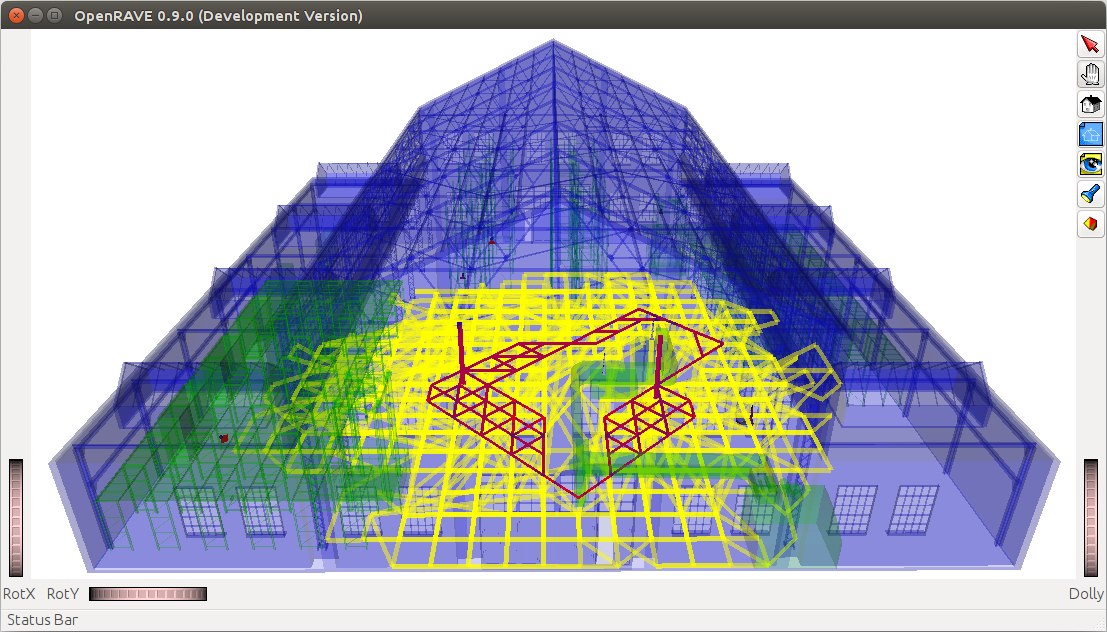}};
        \node (start) at (-4.9, 2.8) {\texttt{start\_pos}};
        \node (target) at (4.8, 2.8) {\texttt{target\_pos}};
        \draw[line width=0.75pt] (start.south) -- (-1.07,-.05);
        \draw[line width=0.75pt] (target.south) -- (1.2,-.2);
    \end{tikzpicture}
    \caption{Screenshot$^{6}$ of \acs{OpenRAVE} including the entire
    environment, the building is depicted in blue and internal objects/obstacles
    are green, the yellow lines mark the procedure (visited position) of the
    algorithm, red lines mark the resulting paths.}
    \label{fig:path}
\end{figure}

The result of the path search is depicted in \fig{fig:path}, it shows an
external view on the situation that was presented at the main page. Although not
directly visible, the red lines actually represent the $1000$ possible paths
that were identified by the algorithm. It has to be note, that the combinatorial
amount of possible paths that require less than $20$ steps exceeds $10^{7}$ (we
tried it). As mentioned, the applied graph traversing algorithm identifies
possible paths according to their length, starting with the shortest ones.

\begin{lstlisting} %
[label=lis:openrave, %
firstnumber=1, %
keywords={SELECT, FROM, WHERE, START, WITH, CONNECT, BY, STOP,
MEMORIZE,MAXIMUM, AS},%
keywordstyle=\color{red}\ttfamily\scriptsize, %
caption={Robotic path search in \acs{OpenRAVE}$^6$.}]
 robot      = "YouBot";
 start_pos  = position(robot);
 target_pos = [10.0, 9.0];
 
 directions = [[0, 1],[0,-1],[1,-1],[-1,-1],[1, 0],[-1, 0],[-1, 1],[1, 1]];

 SELECT (this + cur_pos) FROM directions
 WHERE  target_pos == move(robot, this + cur_pos)

 START WITH cur_pos = start_pos, level = 1
 CONNECT BY MEMORIZE 20 MAXIMUM 1000
            cur_pos = move(robot, cur_pos + this),
            level = level+1
 STOP WITH  target_pos == move(robot, this+cur_pos) OR
            distance(target_pos, this+cur_pos) > 0.5 * (20-level) OR
            IF(checkCollision(robot);
               playSound("/usr/share/sounds/ubuntu/stereo/bell.ogg"),
               True) # ; else is not required here ...
 AS list;
\end{lstlisting}

%% file: sec/outlook.tex
\section{Summary \& Outlook}
\label{sec:outlook}

We have introduced \textsc{SelectScript}, a new declarative embedded
domain-specific language (\acs{DSL}), and shown that it can run as an
intermediate layer on top of different discrete simulation environments and
world models hiding the discrepancies of different \acsp{API}. In addition, we
have demonstrated that accessing information with an \acs{SQL}-like syntax
provides a more ``natural'' and elegant mean than extracting the same
information the old fashioned way, \ie\ looping through the \acs{API}.

The focus of this work lays on the extension of the DSL to better support
reasoning and to hide the complexity of applying different strategies and
methods. Using hierarchical queries the language can be applied stand alone to
solve classical problems, such as the Towers of Hanoi, but also to solve
problems in complex 3D environments, whereby constraints are defined by the
configuration of the applied model, \eg\ walls, obstacles, the geometry and
configuration of the robots, etc.

\textsc{SelectScript} is currently at an early stage, we are exploring different
methods, constructs, and ideas, such as including additional search methods by
applying heuristics or integrating probabilistic aspects to be able to query for
the most likely results. Early development also shows that it is possible to
automatically parallelize the result generation. Although relatively small, our
currently developed interpreter in pure Python is relatively slow. Thus, our
main focus lays in the port of the language to C and in devising a more
efficient implementation. By doing so, we hope that it will be easier to make
\textsc{SelectScript} applicable to other host programming languages.

\textsc{SelectScript} is part of a larger research effort, whereby we try to
make all information origination from an ecology of smart systems accessible and
queryable. In \cite{dietrich2014iros} and \cite{dietrich2014cassandra} we had
described a methodology that overlays smart and distributed environments with
some kind of distributed scene graph (by using cloud based techniques). This
enables us to organize and access these environments in a way that world models
(for a certain area) on the basis of \acs{OpenRAVE} can be reconstructed.
The next step in our research will cover the area of reasoning on functions.
If you look at the \ac{ROS}, it offers a way to access a multitude of
transformation, filter, aggregation, fusion, interpretation functionality. But
how do we use it? Imperatively! Cynically it can be said that we are still
programming assembly, but on far more complex machines. Why not ``simply''
solving this in a declarative way. Reasoning about the sequential application of
``functions'' for data transformation is probably quite similar to the
identification of action sequences. We know what systems are available (their
configuration/pose, data formats, etc.) $[$start configuration$]$, we have this
huge amount of \acs{ROS} functionality (with known input and output formats)
$[$rules$]$ and we know what kind of information/representation we want to have
(as well as a semantic to describe it) $[$goal$]$.